\documentclass[preprint]{aastex}

\begin{document}

\title{WZ Cygni: a Marginal Contact Binary in a Triple System?}
\author{Jae Woo Lee, Seung-Lee Kim, Chung-Uk Lee, Ho-Il Kim, Jang-Ho Park, and Tobias Cornelius Hinse}
\affil{Korea Astronomy and Space Science Institute, Daejeon 305-348, Korea}
\email{jwlee@kasi.re.kr, slkim@kasi.re.kr, leecu@kasi.re.kr, hikim@kasi.re.kr, pooh107162@kasi.re.kr, tobiash@kasi.re.kr}

\begin{abstract}
We present new multiband CCD photometry for WZ Cyg made on 22 nights in two observing seasons of 2007 and 2008. 
Our light-curve synthesis indicates that the system is in poor thermal contact with a fill-out factor of 4.8 \% and 
a temperature difference of 1447 K.  Including our 40 timing measurements, a total of 371 times of minimum light spanning 
more than 112 yr were used for a period study. Detailed analysis of the $O$--$C$ diagram showed that the orbital period 
has varied by a combination with an upward parabola and a sinusoid. The upward parabola means the continuous period increase
and indicates that some stellar masses are thermally transferred from the less to the more massive primary star 
at a rate of about 5.80$\times$10$^{-8}$ M$_\odot$ yr$^{-1}$.  The sinusoidal variation with a period of 47.9 yr and 
a semi-amplitude of 0.008 d can be interpreted most likely as the light-travel-time effect due to the existence of 
a low-mass M-type tertiary companion with a projected mass of $M_3 \sin i_3$=0.26 M$_\odot$.  We examined the evolutionary status 
of WZ Cyg from the absolute dimensions of the eclipsing pair. It belongs to the marginal contact binary systems before 
broken-contact phase, consisting of a massive primary star with spectral type of F4 and a secondary with the type K1.
\end{abstract}

{}

\section{INTRODUCTION}

According to the thermal relaxation oscillation (TRO) theory (Lucy 1976; Lucy \& Wilson 1979), contact binaries oscillate 
between contact and non-contact states on the thermal time-scales ($\sim$$10^7$ yr) because of non-thermal equilibrium.
The systems in good thermal contact show typical W UMa-type light curves with nearly equal eclipse depths, while those 
in broken-contact state display significant temperature differences between the binary components and minima of unequal depth.
Such oscillations are normally accompanied by a thermal mass transfer between both components, and hence cause 
observable period changes. In a state evolving from contact to non-contact phases, mass moves from the less massive component 
toward the more massive star, which is opposite to the energy flow, and an orbital period increase occurs. Since the direction 
of mass transfer can be directly inferred from the observed period change, orbital period studies may provide a significant clue 
to understand the structure and evolutionary state of this kind of system. Such examples are the short-period binaries CN And 
(van Hamme et al. 2001; Lee \& Lee 2006) and V432 Per (Lee et al. 2008b; Odell et al. 2009), which showed 
the secular period decrease and increase, in the same order, caused by mass transfer. The former is a semi-detached system 
with the more massive star filling its limiting lobe and the less massive star very close to lobe-filling, and the latter is 
a detached system in which both components almost fill their inner Roche lobe. The period changes and Roche configurations 
indicate the two systems may be in a broken or marginal contact stage within parameter uncertainties.

WZ Cyg ($\rm BD+38^{o}4262$, 2MASS J20530677+3849406; F4 V) was announced as an eclipsing binary with a short orbital period 
of 0.5845 d (cf. Shapley 1913) and has been studied by several investigators. Shaw (1994) classified the system as 
a near-contact binary (hereafter NCB) whose subclass is unknown.  Rovithis et al. (1999) published 
the first photoelectric light curves and analyzed these.  They suggested that the binary star is a contact system 
of $\beta$ Lyr-type with a photometric mass ratio of $q$=0.54.  Most recently, Siwak et al. (2010) computed 
the binary parameters from their CCD observations and the unpublished radial-velocity curves obtained by Waldemar Ogloza 
and Slavek M. Rucinski at the David Dunlap Observatory. The results indicate that WZ Cyg has a near-contact configuration
with a mass ratio of $q$=0.63, an orbital inclination of $i$=83$^\circ$.50, and a temperature difference of $\Delta T$=1598 K
between the components.

A period study of the system was carried out by Rovithis et al. (1996, 1999), who reported that the orbital period has been 
increasing.  Up to now, eclipse timings have been obtained assiduously by numerous observers and the observation history 
is long enough to investigate the orbital period change. In this paper, we present and analyze new multiband light curves 
and suggest that WZ Cyg is a marginal contact binary in a triple system.

\section{CCD PHOTOMETRIC OBSERVATIONS}

New photometry of WZ Cyg was performed on 22 nights from 2007 October 26 through 2008 October 20 in order to obtain
multicolor light curves and to look for possible flare-like phenomena. The observations were taken with 
a SITe 2K CCD camera and a $BVRI$ filter set attached to the 61-cm reflector at Sobaeksan Optical Astronomy Observatory 
(SOAO) in Korea. The instrument and reduction method are the same as those described by Lee et al. (2007).  
Since an image field-of-view (FOV) was large enough to observe a few tens of nearby stars simultaneously, we monitored 
them along with the program target. As in the method described by Lee et al. (2010), we made an artificial reference source 
from several stars on the CCD frame and examined candidate comparison stars.  Among a few possible candidates, we chose 
TYC 3167-2078-1 (2MASS J20524841+3850375; $V\rm_T$=$+$11.04, $(B-V)\rm_T$=$+$0.57) as an optimal comparison star.

A total of 3194 individual observations was obtained in the four bandpasses (812 in $B$, 799 in $V$, 791 in $R$, 
and 792 in $I$) and a sample of them is listed in Table 1. The natural-system light curves of WZ Cyg are plotted 
in the upper panel of Figure 1 as differential magnitudes {\it versus} orbital phases, where the open circles and plus symbols 
are the individual measures of the 2007 and 2008 seasons, respectively. The differences ('07$-$'08) between the two seasons 
are plotted in the lower panel.  Although we show no figure illustrating variability of the color indices, 
their phase-locked variations are large and in the expected senses. 

In addition to these complete light curves, four primary eclipse timings were obtained in both 2009 and 2010 with 
the same telescope and filters but, replacing the SITe 2K CCD camera, with an electronically cooled FLI IMG4301E CCD camera .
The new CCD chip has 2084$\times$2084 pixels, a pixel size of 24 $\mu$m, and the FOV of about 20$\arcmin$.9$\times$20$\arcmin.9$. 
TYC 3167-1967-1 also served as the comparison star.

\section{LIGHT-CURVE SYNTHESIS AND ABSOLUTE DIMENSIONS}

Our observations for WZ Cyg show a typical light curve of a $\beta$ Lyr-type eclipsing binary (Rovithis et al. 1999; 
Siwak et al. 2010).  Rovithis et al. (1999) reported three flashes made on the 28-29th of June 1993 and suggested that 
these are related to a magnetic activity on the secondary component. However, the other datasets (their observations in 1994, 
Siwak et al. 2010, our data) did not indicate any peculiar light variations.  As shown in the bottom panel of Figure 1, 
mean brightness differences between the 2007 and 2008 seasons are smaller than the observational error of $\pm$0.01 mag: 
$-0.005\pm$0.020 mag for $B$, $-0.002\pm$0.019 mag for $V$, $+0.003\pm$0.021 mag for $R$, and $-0.002\pm$0.022 mag for $I$, 
respectively.  The SOAO observations did not display the year-to-year light variability above the photometric limit.

In order to derive the binary parameters of the system, we analyzed simultaneously our $BVRI$ light curves in a manner 
similar to that for the near-contact binaries RU UMi (Lee et al. 2008a) and GW Gem (Lee et al. 2009) by using 
the 2003 version of the Wilson-Devinney synthesis code (Wilson \& Devinney 1971). The surface temperature of 
the hotter, more massive primary star was assumed to be $T_{1}$=6530 K, according to its spectral type F4 V given by 
Siwak et al. (2010). The gravity-darkening exponents and the bolometric albedos were initialized at standard values 
($g$=0.32 and $A$=0.5) for stars with convective envelopes. The logarithmic bolometric ($X$, $Y$) and 
monochromatic ($x$, $y$) limb-darkening coefficients were interpolated from the values of van Hamme (1993) and were used 
in concert with the model atmosphere option. 

Our light-curve synthesis has been carried out in two stages. In the first stage, all SOAO observations were examined 
for various modes (i.e., Roche configurations) of the binary code and for a series of models with the mass ratios 
in step of 0.02 between 0.4 and 1.0, so as to understand the geometrical structure of the system and to confirm 
a spectroscopic mass ratio of $q$=0.631$\pm$0.036 (Siwak et al. 2010).  Furthermore, a third light source ($\ell_3$) 
was considered throughout the analyses. This procedure showed acceptable photometric solutions only for contact mode 3 
and indicated a somewhat broad range of 0.56 $\le q \le$ 0.64, where the fill-out factor decreases toward larger mass ratios 
from 8 \% at $q$=0.56 to 5 \% at $q$=0.64. But, the trials for a possible $\ell_3$ failed to achieve convergence. 

In the second stage, the previously determined parameters and the mass ratio from Siwak et al. (2010) were used as 
the initial values. Considering its effective temperature with an accuracy no better than 100$\sim$200 K, the envelopes 
of the primary star should lie close to the lower limit between the radiative and convective atmospheres, so $A_1$ and 
$g_1$ were included as additional free variables. Final results are given in Table 2 together with those of 
Siwak et al. (2010) for comparison and plotted in Figure 2, where, for clarity, individual observations have been 
compiled into 100 mean points using bin widths of 0.01 in phase for each filtered light curve. As seen in the figure, 
the computed light curves describe the SOAO multiband data quite well. 

Our result represents the system as a contact binary with a fill-out factor of 4.8 \% and with a considerably large 
temperature difference of 1447 K. This implies that both component stars are marginally over-contact with respect to 
the inner Roche lobe and the thermal contact between them is poor. But contrarily, Siwak et al. (2010) reported WZ Cyg 
to be a NCB in which the two components are at or near their lobes. From our binary parameters and the spectroscopic orbit 
of Siwak et al. (2010), we obtained the absolute dimensions for the system listed in Table 3, assuming that the temperature 
of each component has an error of 200 K and that the bolometric magnitude of the Sun is $M_{\rm bol}$$_\odot$=+4.73. 
For the absolute visual magnitudes ($M_{\rm V}$), we used the bolometric corrections (BCs) from the scaling 
between $\log T$ and BC recalculated by Torres (2010) from Flower's (1996) table.

\section{ORBITAL PERIOD STUDY}

We determined 15 times of minimum light with the weighted means for the timings in each filter by using the method of 
Kwee \& van Woerden (1956). Twenty-five additional timings were derived by us using the data from the WASP 
(Wide Angle Search for Planets) public archive (Butters et al. 2010). From the database of Kreiner et al. (2001) and 
from more recent literature, 331 timings (33 photographic plate, 213 visual, 35 photographic, 11 photoelectric, 
and 39 CCD) have been added to our measurements.  Our period study is based on a total of 371 times of minimum light 
spanning more than 112 years.  All photoelectric and CCD timings are listed in Table 4, wherein the second and third columns 
give the HJED (Heliocentric Julian Ephemeris Date) timings transformed to the terrestrial time scale (Bastian 2000) and 
their uncertainties, respectively.  Because almost all but the CCD timings were published without error information, 
the following standard deviations were assigned to the timing residuals based on observational method 
for the period analysis of WZ Cyg: $\pm$0.0165 d for photographic plate, $\pm$0.0067 d for visual, $\pm$0.0048 d 
for photographic, and $\pm$0.0029 d for photoelectric minima. Relative weights were then calculated as the inverse squares 
of these values consistent with the errors.

As suggested by Rovithis et al. (1996, 1999), we examined whether the orbital period could be represented by 
a quadratic ephemeris via a continuous period increase but failed to give a satisfactory result. Instead, we found 
that all times of minimum light are best fitted by the combination of an upward parabolic variation and 
a light-travel time (LTT) effect caused by the presence of a third body in the system, 
namely a quadratic {\it plus} LTT ephemeris:
\begin{eqnarray*}
C = T_0 + PE + AE^2 + \tau_{3},
\end{eqnarray*}
where $\tau_{3}$ is the LTT due to a third body (Irwin 1952, 1959) and introduces additional five parameters ($a_{12}\sin i_3$, 
$e$, $\omega$, $n$, $T$). Here, $a_{12}\sin i_3$, $e$, and $\omega$ are the orbital parameters of the eclipsing pair 
around the mass center of the triple system. The parameters $n$ and $T$ denote Keplerian mean motion of the mass center 
of the eclipsing pair and the epoch of its periastron passage, respectively. The Levenberg-Marquart algorithm 
(Press et al. 1992) was applied to solve for the unknown parameters of the ephemeris. The results are summarized 
in Table 5, together with the third-body masses ($M_3$) calculated for three different inclinations of $i_3$. 
Our absolute dimensions have been used for these and subsequent calculations.

The $O$--$C$ diagram constructed with the linear terms of the equation is plotted in the top panel of Figure 3, 
where the continuous curve and the dashed parabola represent the full contribution and the quadratic term, respectively. 
The middle panel displays the LTT orbit, and the bottom panel the residuals from the complete ephemeris. These appear 
as $O$--$C_{\rm full}$ in the fifth column of Table 4.  As displayed in Figure 3, the quadratic {\it plus} LTT ephemeris 
currently provides a good representation of all the $O$--$C$ residuals. If the third companion is on the main sequence and 
its orbit is coplanar with the eclipsing binary (i.e., $i_3$=83$^\circ$.2), the mass of the object is $M_3$=0.26 M$_\odot$ 
and its radius and temperature are calculated to be $R_3$=0.27 R$_\odot$ and $T_3$=3048 K, respectively, using 
the mass-radius and mass-temperature relations from well-studied eclipsing binaries (Southworth 2009). These correspond 
to a spectral type of about M6 V and a bolometric luminosity of $L_3$=0.006 L$_\odot$ and contribute about 0.1\% to 
the total light of the triple system. So, it will be difficult to detect such a companion from the light-curve analysis and 
spectroscopic observations.

The positive coefficient of the quadratic term in Table 5 indicates a continuous period increase with a rate of 
3.78 $\times$10$^{-8}$ d yr$^{-1}$, which can be explained by a mass transfer from the cool secondary star to 
its more massive primary component because WZ Cyg is a contact binary system. Under the assumption of 
conservative mass transfer, the transfer rate is 5.80$\times$10$^{-8}$ M$_\odot$ yr$^{-1}$. If the secondary star 
transfers its present mass to the primary component on a thermal time scale $\tau_{\rm th}$=$(GM_{2}^2)/(R_{2}L_{2})$, 
then $\tau_{\rm th}$ = 1.95$\times 10^{7}$ yr and mass is transferred to the primary at a rate given roughly by
$M_{\rm 2}/ \tau_{\rm th}$=5.12$\times 10^{-8}$ M$_{\sun}$ yr$^{-1}$.  This value is very close to 
the mass transfer rate calculated from our quadratic term, which means that the mass transfer between the two stars 
can explain the secular period increase of the system satisfactorily.

\section{DISCUSSION}

The periodic oscillation in the $O$--$C$ residuals could be caused by a magnetic activity cycle in the late-type star, 
as was initially proposed by Applegate (1992) and later modified by Lanza et al. (1998). With the periods 
($P_3$) and amplitudes ($K$) listed in Table 5, the model parameters were calculated from the Applegate formulae and 
are listed in Table 6, where the rms luminosity changes ($\Delta m_{\rm rms}$) converted to magnitude scale were obtained 
with equation (4) in the paper of Kim et al. (1997). In the table, the variations of the gravitational quadrupole moment 
($\Delta Q$) are two orders of magnitude smaller than typical values of $10^{51}\sim10^{52}$ for close binaries 
(Lanza \& Rodono 1999). A recent study by Lanza (2006) indicates that the Applegate mechanism is not adequate to explain 
the orbital period modulation of close binary systems with a late-type secondary. Moreover, it is difficult for the model
to produce perfectly smooth and tilted periodic component in the $O$--$C$ variation. Therefore, the cyclical variation 
most likely arises from the LTT effect due to the existence of an unseen third companion star gravitationally bound to 
the eclipsing pair WZ Cyg.  

Our absolute parameters of WZ Cyg are used to study the evolutionary status of the binary system in the mass-radius and 
mass-luminosity diagrams given by Hilditch et al. (1988).  In these diagrams, the primary star lies in the main-sequence band 
between the zero-age main sequence and the terminal-age main sequence (TAMS), while the secondary is slightly beyond TAMS, 
implying that the component is larger and brighter than expected for its mass. The binary system is closer to 
the general pattern of contact binaries, rather than to that of NCBs.  Nonetheless, the location of the secondary in 
the Hertzsprung-Russell diagram appears to be among NCBs, which significantly depends upon the adopted effective temperatures.  
Considering a period increase driven by mass transfer from the secondary to the primary component, WZ Cyg may be in 
a marginal contact stage evolving from contact to non-contact phases as it undergoes TRO.  

The tertiary component in WZ Cyg may have played an important role in the formation of an initial tidal-locked detached 
progenitor of the eclipsing binary by transferring angular momentum via Kozai oscillation (Kozai 1962; Pribulla \& Rucinski 2006) 
or a combination of the Kozai cycle and tidal friction (Fabrycky \& Tremaine 2007). This would cause WZ Cyg to evlove into 
present configuration by angular momentum loss through magnetic braking and ultimately to coalesce into single stars.  
The existence of the third body is consistent with the suggestion of Pribulla \& Rucinski (2006) that most contact binaries 
exist in multiple systems. Future high-precision long-term observations are needed to verify our results for the orbital behavior 
and the evolutionary status of the system.

\acknowledgments{ }

The authors wish to thank Prof. Chun-Hwey Kim for his help using the $O$--$C$ database of eclipsing binaries and Dr. Michal Siwak 
for sending us their photometric data.  We also thank the staff of the Sobaeksan Optical Astronomy Observatory for assistance 
during our observations. This research has made use of the Simbad database maintained at CDS, Strasbourg, France.  We have used 
data from the WASP public archive in this research.  The WASP consortium comprises of the University of Cambridge, Keele University, 
University of Leicester, The Open University, The Queen¡¯s University Belfast, St. Andrews University and the Isaac Newton Group. 
Funding for WASP comes from the consortium universities and from the UK¡¯s Science and Technology Facilities Council.

\clearpage
\begin{figure}
 \includegraphics[]{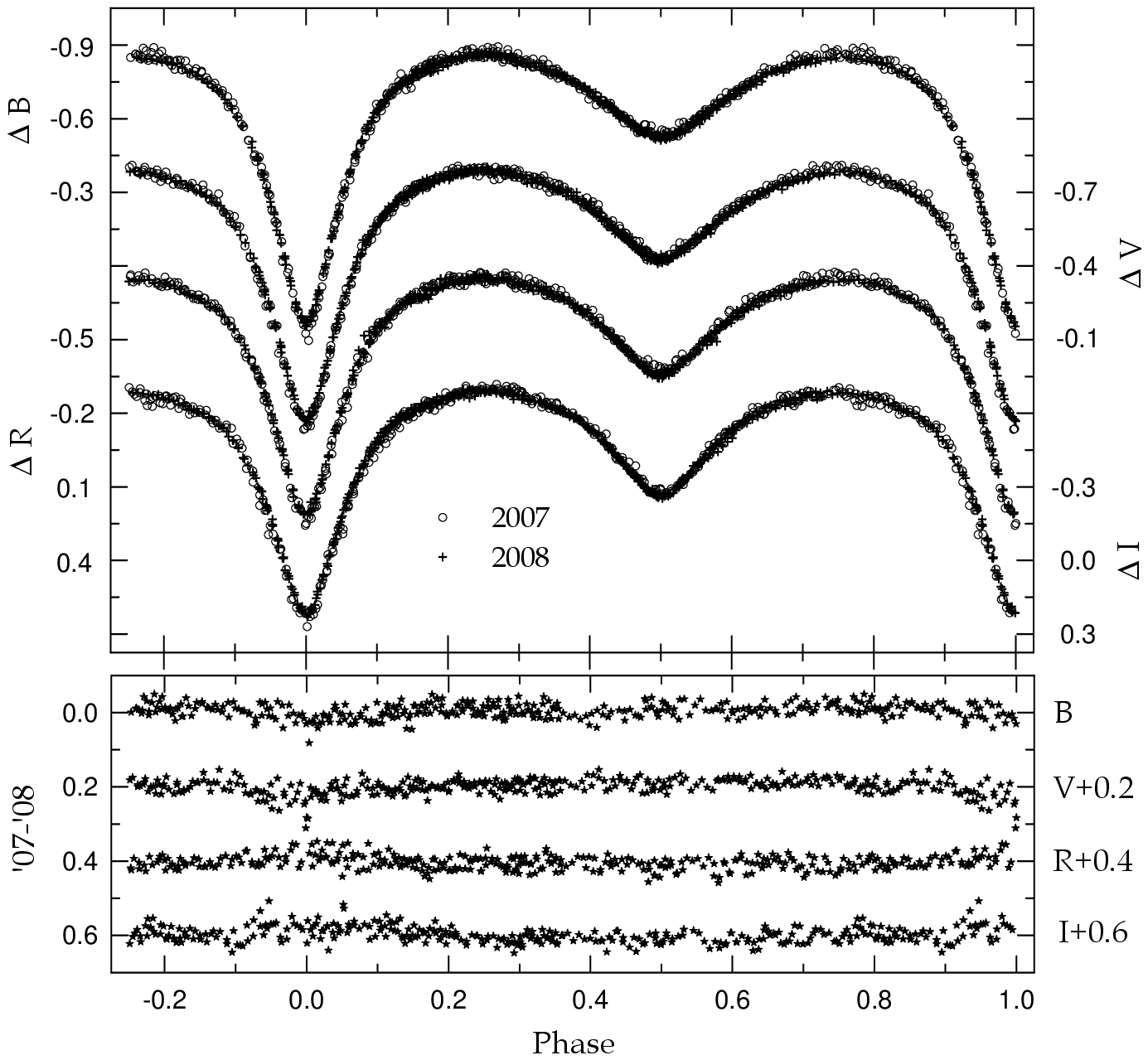}
 \caption{The top panel displays our light curves of WZ Cyg in the $B$, $V$, $R$, and $I$ bandpasses. Because of 
 the high density of the points, many of the 2007 measures cannot be seen individually. The differences between 
 the two seasons are shown in the bottom panel.}
 \label{Fig1}
\end{figure}

\begin{figure}
 \includegraphics[]{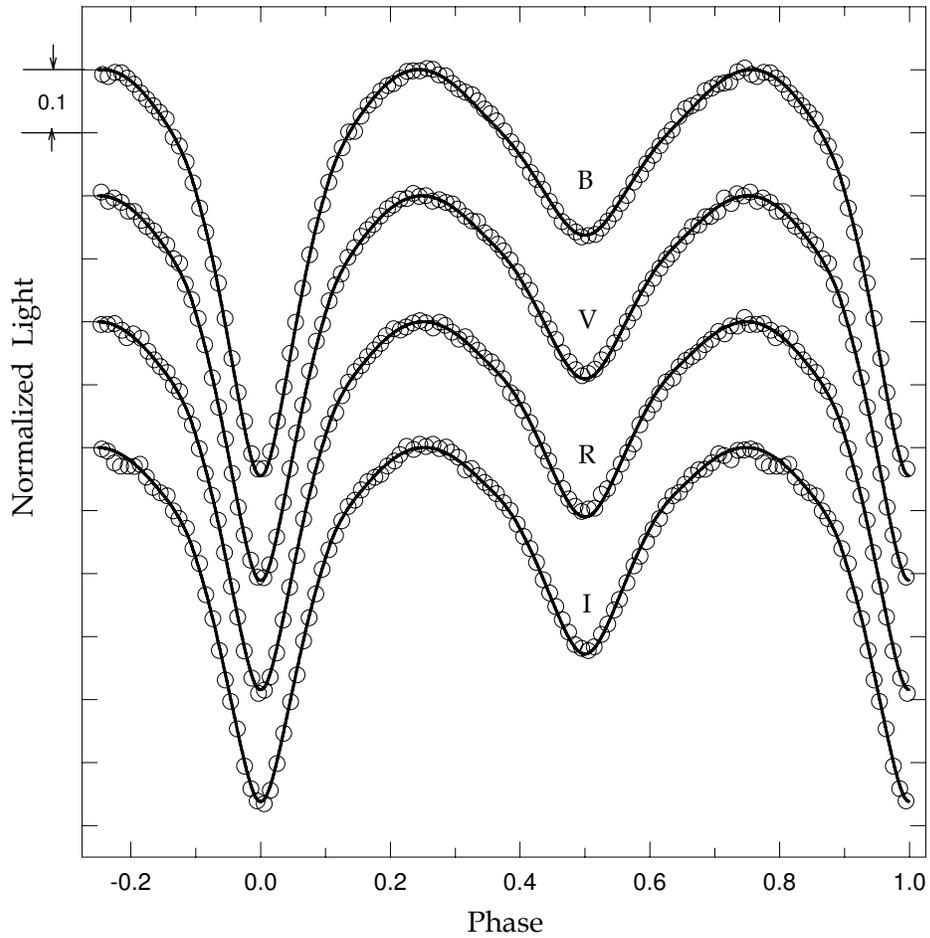}
 \caption{Normalized observations of WZ Cyg with the theoretical light curves obtained by fitting simultaneously
 all SOAO data. For clarity, individual measures have been compiled into 100 mean points using bin widths of 0.01 
 in phase for each filtered light curve. The continuous curves represent the solutions obtained with 
 our model parameters listed in Table 2.}
\label{Fig2}
\end{figure}

\begin{figure}
 \includegraphics[]{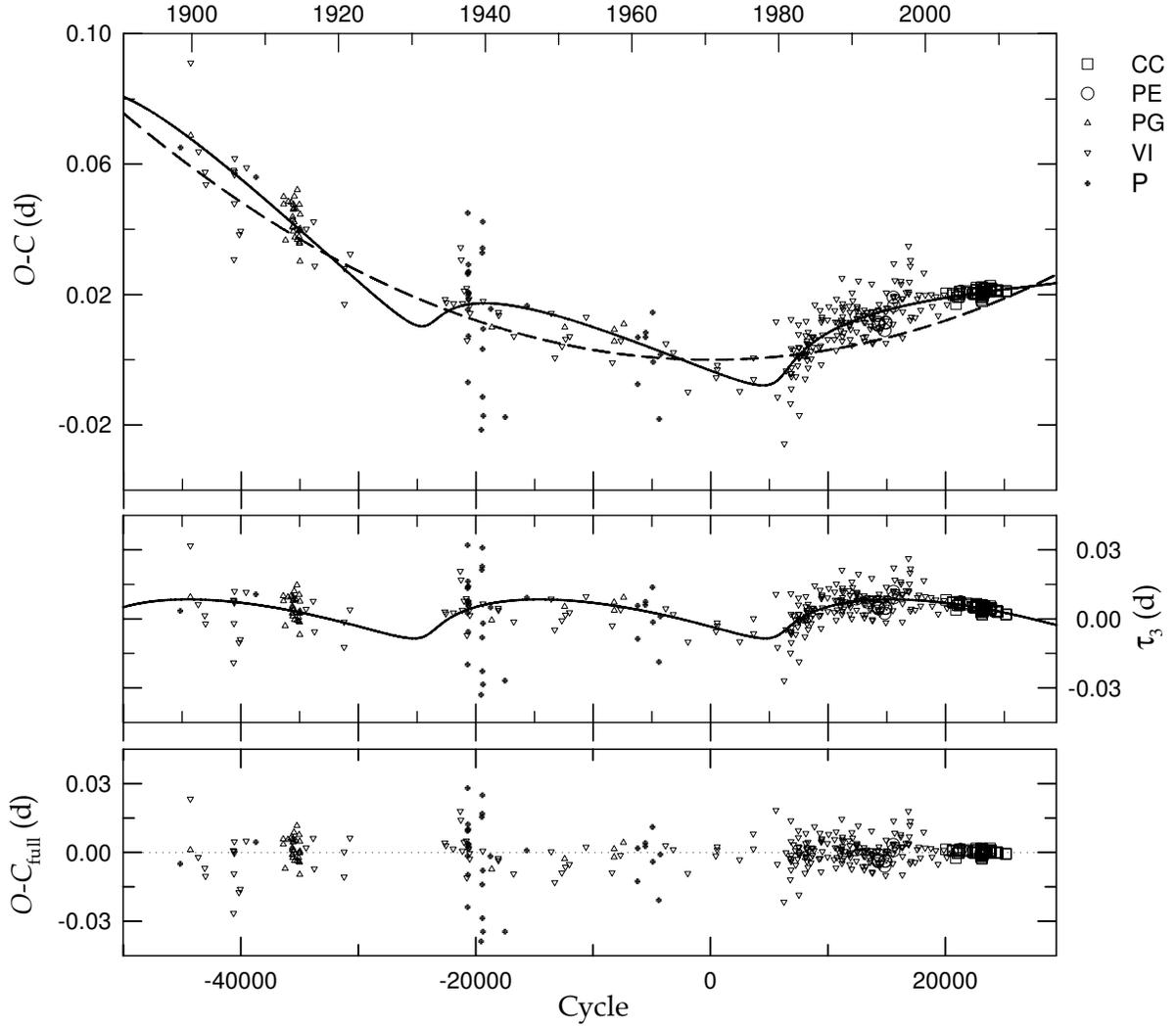}
 \caption{$O$--$C$ diagram of WZ Cyg constructed with the linear terms of the quadratic {\it plus} LTT ephemeris. 
 In the top panel, the continuous curve and the dashed, parabolic one represent the full contribution and 
 the quadratic term of the equation, respectively. The middle panel represents the LTT orbit and the bottom panel 
 the residuals from the complete ephemeris. CC, PE, PG, VI, and P denote CCD, photoelectric, photographic, visual, 
 and photographic plate minima, respectively.}
 \label{Fig3}
\end{figure}

\clearpage
\begin{deluxetable}{crcrcrcr}
\tabletypesize{\small}
\tablewidth{0pt} 
\tablecaption{CCD photometric observations of WZ Cyg.}
\tablehead{
\colhead{HJD} & \colhead{$\Delta B$} & \colhead{HJD} & \colhead{$\Delta V$} & \colhead{HJD} & \colhead{$\Delta R$} & \colhead{HJD} & \colhead{$\Delta I$}
}
\startdata
2,454,399.91834 & $-$0.840  &  2,454,399.91963 & $-$0.768  &  2,454,399.92078 & $-$0.750  &  2454399.92184 & $-$0.696   \\
2,454,399.92314 & $-$0.839  &  2,454,399.92446 & $-$0.782  &  2,454,399.92560 & $-$0.760  &  2454399.92666 & $-$0.707   \\
2,454,399.92795 & $-$0.828  &  2,454,399.92927 & $-$0.774  &  2,454,399.93042 & $-$0.735  &  2454399.93149 & $-$0.674   \\
2,454,399.93282 & $-$0.834  &  2,454,399.93417 & $-$0.767  &  2,454,399.93531 & $-$0.707  &  2454399.93639 & $-$0.666   \\
2,454,399.93773 & $-$0.827  &  2,454,399.93907 & $-$0.755  &  2,454,399.94022 & $-$0.727  &  2454399.94129 & $-$0.650   \\
2,454,399.94262 & $-$0.809  &  2,454,399.94397 & $-$0.761  &  2,454,399.94513 & $-$0.711  &  2454399.94620 & $-$0.658   \\
2,454,399.94753 & $-$0.812  &  2,454,399.94888 & $-$0.757  &  2,454,399.95002 & $-$0.720  &  2454399.95109 & $-$0.638   \\
2,454,399.95242 & $-$0.794  &  2,454,399.95377 & $-$0.717  &  2,454,399.95492 & $-$0.693  &  2454399.95599 & $-$0.620   \\
2,454,399.95732 & $-$0.778  &  2,454,399.95867 & $-$0.716  &  2,454,399.95981 & $-$0.678  &  2454399.96090 & $-$0.612   \\
2,454,399.96223 & $-$0.802  &  2,454,399.96357 & $-$0.728  &  2,454,399.96472 & $-$0.679  &  2454399.96579 & $-$0.610   \\
\enddata
\tablecomments{This table is available in its entirety in machine-readable and Virtual Observatory (VO) forms 
in the online journal. A portion is shown here for guidance regarding its form and content.}
\end{deluxetable}

\begin{deluxetable}{lccccc}
\tablewidth{0pt}
\tablecaption{Binary parameters of WZ Cyg.}
\tablehead{
\colhead{Parameter}             & \multicolumn{2}{c}{Siwak et al. (2010)}   && \multicolumn{2}{c}{This paper}                     \\ [1.0mm] \cline{2-3} \cline{5-6} \\[-2.0ex]
                                & \colhead{Primary} & \colhead{Secondary}   && \colhead{Primary} & \colhead{Secondary}            
}                                                                                                                                 
\startdata                                                                                                                        
$T_0$ (HJD)                     &                       &                   && \multicolumn{2}{c}{2,454,407.938339$\pm$0.000060}  \\
$P$ (d)                         &                       &                   && \multicolumn{2}{c}{0.58446888$\pm$0.00000013}      \\
$q$                             & \multicolumn{2}{c}{0.631}                 && \multicolumn{2}{c}{0.6300$\pm$0.0011}              \\
$i$ (deg)                       & \multicolumn{2}{c}{83.497}                && \multicolumn{2}{c}{83.207$\pm$0.052}               \\
$T$ (K)                         & 6530                  & 4932              && 6530                     & 5083$\pm$9              \\
$\Omega$                        & 3.1200                & 3.1200            && 3.1005$\pm$0.0029        & 3.1005                  \\
$\Omega_{\rm in}$               & \multicolumn{2}{c}{3.1199}                && \multicolumn{2}{c}{3.1182}                         \\
$A$                             & 0.50                  & 0.50              && 0.700$\pm$0.056          & 0.50                    \\
$g$                             & 0.32                  & 0.32              && 0.356$\pm$0.018          & 0.32                    \\
$X$, $Y$                        &                       &                   && 0.639, 0.241             & 0.643, 0.169            \\
$x_{B}$, $y_{B}$                &                       &                   && 0.804, 0.238             & 0.852, 0.001            \\
$x_{V}$, $y_{V}$                &                       &                   && 0.709, 0.277             & 0.796, 0.131            \\
$x_{R}$, $y_{R}$                &                       &                   && 0.615, 0.284             & 0.710, 0.185            \\
$x_{I}$, $y_{I}$                &                       &                   && 0.524, 0.271             & 0.613, 0.201            \\
$L/(L_1+L_2)_{B}$               & 0.8691                & 0.1309            && 0.8852$\pm$0.0019        & 0.1148                  \\
$L/(L_1+L_2)_{V}$               & 0.8453                & 0.1547            && 0.8381$\pm$0.0017        & 0.1619                  \\
$L/(L_1+L_2)_{R}$               & 0.8233                & 0.1767            && 0.8034$\pm$0.0015        & 0.1966                  \\
$L/(L_1+L_2)_{I}$               & 0.7854                & 0.2146            && 0.7737$\pm$0.0013        & 0.2263                  \\
$r$ (pole)                      & 0.3948                & 0.3183            && 0.3976$\pm$0.0005        & 0.3209$\pm$0.0005       \\
$r$ (side)                      & 0.4174                & 0.3327            && 0.4210$\pm$0.0006        & 0.3359$\pm$0.0006       \\
$r$ (back)                      & 0.4465                & 0.3649            && 0.4512$\pm$0.0007        & 0.3697$\pm$0.0010       \\
$r$ (volume)$^\dagger$         & 0.4210                & 0.3400            && 0.4250                   & 0.3439                  \\
\enddata
\tablenotetext{\dagger}{Mean volume radius.}
\end{deluxetable}

\begin{deluxetable}{lcc}
\tablewidth{0pt}
\tablecaption{Absolute parameters for WZ Cyg.}
\tablehead{
\colhead{Parameter}    & \colhead{Primary}    & \colhead{Secondary}}
\startdata                                    
$M$ (M$_\odot$)        &  1.59$\pm$0.04       &  1.00$\pm$0.02       \\
$R$ (R$_\odot$)        &  1.72$\pm$0.02       &  1.39$\pm$0.02       \\
$\log$ $g$ (cgs)       &  4.17$\pm$0.02       &  4.15$\pm$0.01       \\
$\rho$ (g cm$^3)$      &  0.44$\pm$0.02       &  0.53$\pm$0.02       \\
$T$ (K)                &  6530$\pm$200        &  5083$\pm$200        \\
$L$ (L$_\odot$)        &  4.80$\pm$0.60       &  1.15$\pm$0.18       \\
$M_{\rm bol}$ (mag)    &  $+$3.03$\pm$0.14    &  $+$4.57$\pm$0.11    \\
BC (mag)               &  $+$0.01             &  $-$0.27             \\
$M_{\rm V}$ (mag)      &  $+$3.02$\pm$0.14    &  $+$4.84$\pm$0.11    \\
\enddata                                                     
\end{deluxetable}

\begin{deluxetable}{lllrrcl}
\tabletypesize{\small} 
\tablewidth{0pt}
\tablecaption{Photoelectric and CCD timings of minimum light for WZ Cyg.}
\tablehead{
\colhead{HJD} & \colhead{HJED} & \colhead{Error} & \colhead{Epoch} & \colhead{$O$--$C_{\rm full}$} & \colhead{Min} & References \\
\colhead{(2,400,000+)} & \colhead{(2,400,000+)} & & & & & }                                                               
\startdata                                                                                     
48,073.4709   &  48,073.47156  &                &  12401.0  &  $+$0.00111  &  I   &  Hanzl(1991)                   \\  
49,163.5002   &  49,163.50088  &                &  14266.0  &  $-$0.00396  &  I   &  Rovithis et al. (1996)        \\  
49,164.3800   &  49,164.38068  &                &  14267.5  &  $-$0.00086  &  II  &  Rovithis et al. (1996)        \\  
49,168.4685   &  49,168.46918  &                &  14274.5  &  $-$0.00364  &  II  &  Rovithis et al. (1996)        \\  
49,169.3457   &  49,169.34640  &                &  14276.0  &  $-$0.00312  &  I   &  Rovithis et al. (1996)        \\  
49,490.5083   &  49,490.50900  &                &  14825.5  &  $-$0.00605  &  II  &  Rovithis et al. (1996)        \\  
49,529.3778   &  49,529.37850  &                &  14892.0  &  $-$0.00372  &  I   &  Rovithis et al. (1996)        \\  
49,530.5466   &  49,530.54730  &                &  14894.0  &  $-$0.00386  &  I   &  Rovithis et al. (1996)        \\  
49,917.4720   &  49,917.47271  &  $\pm$0.00040  &  15556.0  &  $+$0.00335  &  I   &  Albayrak et al. (2000)        \\  
49,938.5103   &  49,938.51101  &  $\pm$0.00030  &  15592.0  &  $+$0.00078  &  I   &  Albayrak et al. (2000)        \\  
52,546.4087   &  52,546.40944  &  $\pm$0.00070  &  20054.0  &  $+$0.00132  &  I   &  Agerer \& H\"ubscher (2003)   \\  
52,870.2037   &  52,870.20444  &  $\pm$0.00020  &  20608.0  &  $+$0.00092  &  I   &  Nagai (2004)                  \\  
52,902.3490   &  52,902.34974  &  $\pm$0.00100  &  20663.0  &  $+$0.00047  &  I   &  Br\'at et al. (2007)          \\  
53,040.2806   &  53,040.28134  &  $\pm$0.00080  &  20899.0  &  $-$0.00243  &  I   &  Sobotka (2007)                \\  
53,116.2635   &  53,116.26424  &  $\pm$0.00020  &  21029.0  &  $-$0.00040  &  I   &  Nagai (2005)                  \\  
53,193.4134   &  53,193.41414  &  $\pm$0.00130  &  21161.0  &  $-$0.00030  &  I   &  Br\'at et al. (2007)          \\  
53,202.7655   &  53,202.76624  &  $\pm$0.00050  &  21177.0  &  $+$0.00031  &  I   &  Sobotka (2007)                \\  
53,206.2728   &  53,206.27354  &  $\pm$0.00030  &  21183.0  &  $+$0.00080  &  I   &  Sobotka (2007)                \\  
53,224.3918   &  53,224.39254  &  $\pm$0.00180  &  21214.0  &  $+$0.00129  &  I   &  Br\'at et al. (2007)          \\  
53,259.4598   &  53,259.46054  &  $\pm$0.00130  &  21274.0  &  $+$0.00120  &  I   &  H\"ubscher et al. (2005)      \\  
53,546.4336   &  53,546.43434  &  $\pm$0.00180  &  21765.0  &  $+$0.00112  &  I   &  Br\'at et al. (2007)          \\  
53,612.4780   &  53,612.47874  &  $\pm$0.00380  &  21878.0  &  $+$0.00061  &  I   &  H\"ubscher et al. (2006)      \\  
53,887.1776   &  53,887.17835  &  $\pm$0.00020  &  22348.0  &  $+$0.00019  &  I   &  Nagai (2007)                  \\  
53,920.4919   &  53,920.49265  &  $\pm$0.00010  &  22405.0  &  $-$0.00020  &  I   &  H\"ubscher (2007)             \\  
53,985.3679   &  53,985.36865  &  $\pm$0.00010  &  22516.0  &  $-$0.00016  &  I   &  Do\u{g}ru et al. (2007)       \\  
54,003.4867   &  54,003.48745  &  $\pm$0.00010  &  22547.0  &  $+$0.00012  &  I   &  H\"ubscher (2007)             \\  
54,018.39034  &  54,018.39109  &  $\pm$0.00020  &  22572.5  &  $-$0.00017  &  II  &  Siwak et al. (2010)           \\  
54,019.26728  &  54,019.26803  &  $\pm$0.00010  &  22574.0  &  $+$0.00006  &  I   &  Siwak et al. (2010)           \\  
54,020.43612  &  54,020.43687  &  $\pm$0.00005  &  22576.0  &  $-$0.00003  &  I   &  Siwak et al. (2010)           \\  
54,279.64626  &  54279.64701   &  $\pm$0.00063  &  23019.5  &  $-$0.00151  &  II  &  This paper (WASP)             \\  
54,282.57021  &  54282.57096   &  $\pm$0.00036  &  23024.5  &  $+$0.00009  &  II  &  This paper (WASP)             \\  
54,284.61529  &  54284.61604   &  $\pm$0.00026  &  23028.0  &  $-$0.00046  &  I   &  This paper (WASP)             \\  
54,286.66135  &  54286.66210   &  $\pm$0.00040  &  23031.5  &  $-$0.00004  &  II  &  This paper (WASP)             \\  
54,287.53895  &  54287.53970   &  $\pm$0.00027  &  23033.0  &  $+$0.00086  &  I   &  This paper (WASP)             \\  
54,288.70734  &  54288.70809   &  $\pm$0.00026  &  23035.0  &  $+$0.00031  &  I   &  This paper (WASP)             \\  
54,289.58525  &  54289.58600   &  $\pm$0.00038  &  23036.5  &  $+$0.00152  &  II  &  This paper (WASP)             \\  
54,291.62883  &  54291.62958   &  $\pm$0.00016  &  23040.0  &  $-$0.00054  &  I   &  This paper (WASP)             \\  
54,292.50708  &  54292.50783   &  $\pm$0.00035  &  23041.5  &  $+$0.00101  &  II  &  This paper (WASP)             \\  
54,298.64267  &  54298.64342   &  $\pm$0.00010  &  23052.0  &  $-$0.00032  &  I   &  This paper (WASP)             \\  
54,304.48665  &  54304.48740   &  $\pm$0.00024  &  23062.0  &  $-$0.00102  &  I   &  This paper (WASP)             \\  
54,306.53106  &  54306.53181   &  $\pm$0.00067  &  23065.5  &  $-$0.00225  &  II  &  This paper (WASP)             \\  
54,308.57840  &  54308.57915   &  $\pm$0.00024  &  23069.0  &  $-$0.00055  &  I   &  This paper (WASP)             \\  
54,326.1127   &  54326.11345   &  $\pm$0.00010  &  23099.0  &  $-$0.00029  &  I   &  Nagai (2008)                  \\  
54,333.42045  &  54333.42120   &  $\pm$0.00017  &  23111.5  &  $+$0.00161  &  II  &  This paper (WASP)             \\  
54,335.46381  &  54335.46456   &  $\pm$0.00015  &  23115.0  &  $-$0.00067  &  I   &  This paper (WASP)             \\  
54,337.51039  &  54337.51114   &  $\pm$0.00040  &  23118.5  &  $+$0.00027  &  II  &  This paper (WASP)             \\  
54,338.38632  &  54338.38707   &  $\pm$0.00021  &  23120.0  &  $-$0.00050  &  I   &  This paper (WASP)             \\  
54,339.55539  &  54339.55614   &  $\pm$0.00020  &  23122.0  &  $-$0.00037  &  I   &  This paper (WASP)             \\  
54,340.42969  &  54340.43044   &  $\pm$0.00049  &  23123.5  &  $-$0.00277  &  II  &  This paper (WASP)             \\  
54,344.52371  &  54344.52446   &  $\pm$0.00047  &  23130.5  &  $-$0.00003  &  II  &  This paper (WASP)             \\  
54,346.56912  &  54346.56987   &  $\pm$0.00021  &  23134.0  &  $-$0.00026  &  I   &  This paper (WASP)             \\  
54,364.39486  &  54364.39561   &  $\pm$0.00044  &  23164.5  &  $-$0.00079  &  II  &  This paper (WASP)             \\  
54,371.40781  &  54371.40856   &  $\pm$0.00042  &  23176.5  &  $-$0.00146  &  II  &  This paper (WASP)             \\  
54,373.45455  &  54373.45530   &  $\pm$0.00021  &  23180.0  &  $-$0.00036  &  I   &  This paper (WASP)             \\  
54,395.37267  &  54395.37342   &  $\pm$0.00048  &  23217.5  &  $+$0.00020  &  II  &  This paper (WASP)             \\  
54,397.41793  &  54397.41868   &  $\pm$0.00022  &  23221.0  &  $-$0.00017  &  I   &  This paper (WASP)             \\  
54,400.04833  &  54,400.04908  &  $\pm$0.00059  &  23225.5  &  $+$0.00012  &  II  &  This paper (SOAO)             \\  
54,407.93824  &  54,407.93899  &  $\pm$0.00028  &  23239.0  &  $-$0.00029  &  I   &  This paper (SOAO)             \\  
54,426.05679  &  54,426.05754  &  $\pm$0.00024  &  23270.0  &  $-$0.00025  &  I   &  This paper (SOAO)             \\  
54,433.3627   &  54,433.36345  &  $\pm$0.00030  &  23282.5  &  $-$0.00019  &  II  &  Br\'at et al. (2008)          \\  
54,455.2815   &  54,455.28225  &  $\pm$0.00010  &  23320.0  &  $+$0.00105  &  I   &  H\"ubscher et al. (2009)      \\  
54,712.15401  &  54,712.15476  &  $\pm$0.00020  &  23759.5  &  $-$0.00018  &  II  &  This paper (SOAO)             \\  
54,716.24572  &  54,716.24647  &  $\pm$0.00033  &  23766.5  &  $+$0.00025  &  II  &  This paper (SOAO)             \\  
54,741.96376  &  54,741.96451  &  $\pm$0.00038  &  23810.5  &  $+$0.00170  &  II  &  This paper (SOAO)             \\  
54,748.97644  &  54,748.97719  &  $\pm$0.00043  &  23822.5  &  $+$0.00076  &  II  &  This paper (SOAO)             \\  
54,751.02142  &  54,751.02217  &  $\pm$0.00013  &  23826.0  &  $+$0.00010  &  I   &  This paper (SOAO)             \\  
54,753.94393  &  54,753.94468  &  $\pm$0.00008  &  23831.0  &  $+$0.00027  &  I   &  This paper (SOAO)             \\  
54,755.11277  &  54,755.11352  &  $\pm$0.00008  &  23833.0  &  $+$0.00017  &  I   &  This paper (SOAO)             \\  
54,760.08100  &  54,760.08175  &  $\pm$0.00020  &  23841.5  &  $+$0.00042  &  II  &  This paper (SOAO)             \\  
54,798.3641   &  54,798.36485  &  $\pm$0.00010  &  23907.0  &  $+$0.00086  &  I   &  H\"ubscher et al. (2009)      \\  
55,018.1231   &  55,018.12387  &  $\pm$0.00010  &  24283.0  &  $-$0.00013  &  I   &  Nagai (2009)                  \\  
55,079.4924   &  55,079.49317  &  $\pm$0.00030  &  24388.0  &  $+$0.00002  &  I   &  Erkan et al. (2010)           \\  
55,091.18180  &  55,091.18257  &  $\pm$0.00007  &  24408.0  &  $+$0.00006  &  I   &  This paper (SOAO)             \\  
55,115.14457  &  55,115.14534  &  $\pm$0.00020  &  24449.0  &  $-$0.00036  &  I   &  This paper (SOAO)             \\  
55,526.02520  &  55,526.02597  &  $\pm$0.00013  &  25152.0  &  $-$0.00081  &  I   &  This paper (SOAO)             \\  
55,528.94785  &  55,528.94862  &  $\pm$0.00015  &  25157.0  &  $-$0.00050  &  I   &  This paper (SOAO)             \\  
\enddata
\end{deluxetable}

\begin{deluxetable}{lcc}
\tablewidth{0pt}
\tablecaption{Parameters for the quadratic {\it plus} LTT ephemeris of WZ Cyg. }
\tablehead{
\colhead{Parameter}               &  \colhead{Values}            &  \colhead{Unit}
}
\startdata                        
$T_0$                             &  2,440,825.47487$\pm$0.00061          &  HJED                  \\
$P$                               &  0.584467647$\pm$0.000000027          &  d                     \\
$A$                               &  +(3.02$\pm$0.11)$\times 10^{-11}$    &  d                     \\
$a_{12}\sin i_{3}$                &  1.72$\pm$0.21                        &  AU                    \\
$\omega$                          &  320.0$\pm$4.1                        &  deg                   \\
$e$                               &  0.67$\pm$0.19                        &                        \\
$n  $                             &  0.02058$\pm$0.00078                  &  deg d$^{-1}$          \\
$T$                               &  2,444,328$\pm$379                    &  HJED                  \\
$P_{3}$                           &  47.9$\pm$1.8                         &  yr                    \\
$K$                               &  0.0085$\pm$0.0011                    &  d                     \\
$f(M_{3})$                        &  0.00220$\pm$0.00029                  &  $M_\odot$             \\
$M_3$ ($i_{3}$=90 deg)$^\dagger$  &  0.26                                 &  $M_\odot$             \\
$M_3$ ($i_{3}$=60 deg)$^\dagger$  &  0.31                                 &  $M_\odot$             \\
$M_3$ ($i_{3}$=30 deg)$^\dagger$  &  0.56                                 &  $M_\odot$             \\
$dP$/$dt$                         &  +(3.78$\pm$0.14)$\times 10^{-8}$     &  d yr$^{-1}$           \\
$dM_2$/$dt$                       &  $-$5.80$\times 10^{-8}$              &  $M_\odot$ yr$^{-1}$   \\
\enddata
\tablenotetext{\dagger}{Hypothetical third-body masses for different values of $i_{3}$.}
\end{deluxetable}

\clearpage
\begin{deluxetable}{lccc}
\tablewidth{0pt}
\tablecaption{Applegate parameters for possible magnetic activity of WZ Cyg.}
\tablehead{
\colhead{Parameter}       & \colhead{Primary}      & \colhead{Secondary}     & \colhead{Unit}
}
\startdata
$\Delta P$                & 0.1523                 &  0.1523                 &  s                   \\
$\Delta P/P$              & $3.02\times10^{-6}$    &  $3.02\times10^{-6}$    &                      \\
$\Delta Q$                & ${8.38\times10^{49}}$  &  ${5.27\times10^{49}}$  &  g cm$^2$            \\
$\Delta J$                & ${2.49\times10^{47}}$  &  ${1.86\times10^{47}}$  &  g cm$^{2}$ s$^{-1}$ \\
$I_{\rm s}$               & ${3.02\times10^{54}}$  &  ${1.24\times10^{54}}$  &  g cm$^{2}$          \\
$\Delta \Omega$           & ${8.22\times10^{-8}}$  &  ${1.50\times10^{-7}}$  &  s$^{-1}$            \\
$\Delta \Omega / \Omega$  & ${6.61\times10^{-4}}$  &  ${1.21\times10^{-3}}$  &                      \\
$\Delta E$                & ${4.09\times10^{40}}$  &  ${5.59\times10^{40}}$  &  erg                 \\
$\Delta L_{\rm rms}$      & ${8.50\times10^{31}}$  &  ${1.16\times10^{32}}$  &  erg s$^{-1}$        \\
                          & 0.0218                 &  0.0298                 &  $L_\odot$           \\
                          & 0.0045                 &  0.0259                 &  $L_{1,2}$           \\
$\Delta m_{\rm rms}$      & $\pm$0.0040            &  $\pm$0.0054            &  mag                 \\
$B$                       & 4,350                  &  5,183                  &  G                   
\enddata
\end{deluxetable}


\newpage
\begin{thebibliography}{}
\bibitem[Agerer \& Hubscher(2003)]{agerer2003} Agerer, F., \& H\"ubscher, J. 2003, Inf. Bull. Var. Stars, 5484, 1
\bibitem[Albayrak et al(2000)]{albayrak2000} Albayrak, B., M\"uyesseroglu, Z., \& \"Ozdemir, S. 2000, Inf. Bull. Variable Stars, 4941, 1
\bibitem[Applegate(1992)]{applegate1999} Applegate, J. H. 1992, ApJ, 385, 621
\bibitem[Bastian(2000)]{bastian2000} Bastian, U. 2000, Inf. Bull. Variable Stars, 4822, 1
\bibitem[Brat et al(2007)]{brat2007} Br\'at, L., Zejda, M., \& Svoboda, P. 2007, Open Eur. J. Var. Stars, 74, 1
\bibitem[Brat et al(2008)]{brat2008} Br\'at, L., et al. 2008, Open Eur. J. Var. Stars, 94, 1
\bibitem[Butters et al(2010)]{butters2010} Butters, O. W., et al. 2010, A\&A, 520, L10 
\bibitem[Dogru et al(2007)]{dogru2007} Do\u{g}ru, S. S., D\"onmez, A., T\"uys\"uz, M., Do\u{g}ru, D., \"Ozkarde\c s, B., Soydugan, E., \& Soydugan, F. 2007, Inf. Bull. Variable Stars, 5746, 1
\bibitem[Erkan et al(2010)]{erkan2010} Erkan, N., Erdem, A., Akin, T., Alicavus, F., \& Soydugan, F. 2010, Inf. Bull. Variable Stars, 5924, 1
\bibitem[Flower(1996)]{flower1996} Flower, P. J. 1996, ApJ, 469, 355
\bibitem[Fabrycky \& Tremaine(2007)]{fabrycky2007} Fabrycky, D., \& Tremaine, S. 2007, ApJ, 669, 1298
\bibitem[Hanzl(1991)]{hanzl1991} Hanzl, D. 1991, Inf. Bull. Variable Stars, 3615, 1
\bibitem[Hilditch et al(1988)]{hilditch1988} Hilditch, R. W., King, D. J., \& McFarlane, T. M. 1988, MNRAS, 231, 341
\bibitem[H\"ubscher(2007)]{hubscher2007} H\"ubscher, J. 2007, Inf. Bull. Var. Stars, 5802, 1
\bibitem[H\"ubscher et al(2005)]{hubscher2005} H\"ubscher, J., Paschke, A., \& Walter, F. 2005, Inf. Bull. Var. Stars, 5657, 1
\bibitem[H\"ubscher et al(2006)]{hubscher2006} H\"ubscher, J., Paschke, A., \& Walter, F. 2006, Inf. Bull. Var. Stars, 5731, 1
\bibitem[H\"ubscher et al(2009)]{hubscher2009} H\"ubscher, J., Steinbach, H.-M., \& Walter, F. 2009, Inf. Bull. Var. Stars, 5874, 1
\bibitem[Irwin(1952)]{irwin1952} Irwin, J. B. 1952, ApJ, 116, 211
\bibitem[Irwin(1959)]{irwin1959} Irwin, J. B. 1959, AJ, 64, 149
\bibitem[Kim et al(1997)]{kim1997} Kim, C.-H., Jeong, J. H., Demircan, O., M\"{u}yessero\u{g}lu, Z., \& Budding, E. 1997, AJ, 114, 2753
\bibitem[Kozai(1962)]{kozai1962} Kozai, Y. 1962, AJ, 67, 579
\bibitem[Kreiner et al(2001)]{kreiner2001} Kreiner, J. M., Kim, C.-H., \& Nha, I.-S. 2001, An Atlas of $O$--$C$ Diagrams of Eclipsing Binary Stars (Krakow: Wydawn. Nauk. Akad. Pedagogicznej)
\bibitem[Kwee \& van Woerden(1956)]{kwee1956} Kwee, K. K., \& van Woerden, H. 1956, Bull. Astron. Inst. Neth., 12, 327
\bibitem[Lanza(2006)]{lanza2006} Lanza, A. F. 2006, MNRAS, 369, 1773
\bibitem[Lanza \& Rodono(1999)]{lanza1999} Lanza, A. F., \& Rodono, M. 1999, A\&A, 349, 887
\bibitem[Lanza, Rodono \& Rosner(1998)]{lanza1998} Lanza, A. F., Rodono, M., \& Rosner, R. 1998, MNRAS, 296, 893
\bibitem[Lee \& Lee(2006)]{lee2006} Lee, C.-U., \& Lee, J. W. 2006, J. Korean Astron. Soc., 39, 25
\bibitem[Lee et al(2007)]{lee2007} Lee, J. W., Kim, C.-H., \& Koch, R. H. 2007, MNRAS, 379, 1665
\bibitem[Lee et al(2008a)]{lee2008a} Lee, J. W., Kim, C.-H., Kim, S.-L., Lee, C.-U., Han, W., \& Koch, R. H. 2008a, PASP, 120, 720
\bibitem[Lee et al(2008b)]{lee2008b} Lee, J. W., Youn, J.-H., Kim, C.-H., Lee, C.-U., \& Kim, H.-I. 2008b, AJ, 135, 1523
\bibitem[Lee et al(2009)]{lee2009} Lee, J. W., Kim, S.-L., Lee, C.-U., \& Youn, J.-H. 2009, PASP, 121, 1366
\bibitem[Lee et al(2010)]{lee2010} Lee, J. W., et al. 2010, AJ, 139, 898
\bibitem[Lucy(1976)]{lucy1976} Lucy, L. B. 1976, ApJ, 205, 208
\bibitem[Lucy \& Wilson(1979)]{lucy1979} Lucy, L. B., \& Wilson, R. E. 1979, ApJ, 231, 502
\bibitem[Nagai(2004)]{nagai2004} Nagai, K. 2004, VSOLJ Var. Star Bull., 42, 2
\bibitem[Nagai(2005)]{nagai2005} Nagai, K. 2005, VSOLJ Var. Star Bull., 43, 3
\bibitem[Nagai(2007)]{nagai2007} Nagai, K. 2007, VSOLJ Var. Star Bull., 45, 3
\bibitem[Nagai(2008)]{nagai2008} Nagai, K. 2008, VSOLJ Var. Star Bull., 46, 3
\bibitem[Nagai(2009)]{nagai2009} Nagai, K. 2009, VSOLJ Var. Star Bull., 50, 3
\bibitem[Odell et al(2009)]{odell2009} Odell, A. P., Eaton, J. A., \& L\'opez-Cruz, O. 2009, MNRAS, 400, 2085
\bibitem[Press et al(1992)]{press1992} Press, W. H., Teukolsky, S. A., Vetterling, W. T., \& Flannery, B. P. 1992, Numerical Recipes (Cambridge: Cambridge Univ. Press), Chapter 15
\bibitem[Pribulla \& Rucinski(2006)]{pribulla2006} Pribulla, T., \& Rucinski, S. M. 2006, AJ, 131, 2986
\bibitem[Rovithis et al(1996)]{rovithis1996} Rovithis, P., Rovithis-Livaniou, H., \& Kranidiotis, A. 1996, Inf. Bull. Var. Stars, 4309, 1
\bibitem[Rovithis et al(1999)]{rovithis1999} Rovithis, P., Rovithis-Livaniou, H., Suran, M. D., Fragoulopoulou, E., \& Skopal, A. 1999, A\&A, 348, 184 
\bibitem[Shapley(1913)]{shapley1913} Shapley, H. 1913, ApJ, 38, 158
\bibitem[Shaw(1994)]{shaw1994} Shaw, J. S. 1994, Mem. Soc. Astron. Italiana, 65, 95
\bibitem[Siwak et al(2010)]{siwak2010} Siwak, M., Zola, S., \& Koziel-Wierzbowska, D. 2010, Acta Astron., 60, 305
\bibitem[Sobotka(2007)]{sobotka2007} Sobotka, P. 2007, Inf. Bull. Variable Stars, 5809, 1
\bibitem[Southworth(2009)]{southworth2009} Southworth, J. 2009, MNRAS, 394, 272
\bibitem[Torres(2010)]{torres2010} Torres, G. 2010, AJ, 140, 1158
\bibitem[Wilson \& Devinney(1971)]{wilson1971} Wilson, R. E., \& Devinney, E. J. 1971, ApJ, 166, 605
\bibitem[Van Hamme(1993)]{van1993} Van Hamme, W. 1993, AJ, 106, 209
\bibitem[Van Hamme et al(2001)]{van2001} Van Hamme, W., Samec, R. G., Gothard, N. W., Wilson, R. E., Faulkner, D. R., \& Branly, R. M. 2001, AJ, 122, 3436
\end{thebibliography}
\end{document}